# THE DEVASTING ECONOMIC IMPACT OF *CALLINECTES SAPIDUS* ON THE CLAM FISHING IN THE PO DELTA (ITALY): STRIKING EVIDENCE FROM NOVEL FIELD DATA

Francesco Tiralongo[1,2,3*], Emanuele Mancini[2,4,5], Raffaele Gattelli[6], Costanza Di Pasquale[7], Emanuele Rossetti[8], Riccardo Martellucci[9], Alberto Felici[7], Giorgio Mancinelli[4,5,10], Christian Mulder[1,10**]

[1]Department of Biological, Geological and Environmental Sciences, University of Catania, 95124 Catania, Italy
[2]Ente Fauna Marina Mediterranea, Scientific Organization for Research and Conservation of Marine Biodiversity, 96012 Avola, Italy
[3]National Research Council, Institute of Marine Biological Resources and Biotechnologies, 60125 Ancona, Italy
[4]Department of Biological and Environmental Sciences and Technologies, DiSTeBA, University of Salento, 73100 Lecce, Italy
[5]National Biodiversity Future Center (NBFC), 90100 Palermo, Italy
[6]Department of Biological, Geological and Environmental Sciences, University of Bologna, 40126 Bologna, Italy
[7]School of Biosciences and Veterinary Medicine, University of Camerino, 62024 Matelica, Italy
[8]Consorzio Cooperative Pescatori del Polesine O.P. S.C.Ar.L., 45018 Scardovari, Italy
[9]National Institute of Oceanography and Applied Geophysics (OGS), 34010 Trieste, Italy
[10] CoNISMa, Piazzale Flaminio 9, 00197 Roma, Italy

* francesco.tiralongo@unict.it, ** christian.mulder@unict.it

**Abstract**

Invasive species are a growing threat to marine ecosystems, and the recent proliferation of the Atlantic blue crab (*Callinectes sapidus*) in the Po Delta (Italy) has had significant ecological and economic impacts, particularly on clam farming. This study explores the influence of *C. sapidus* on clam production in the Po Delta, combining biological and ecological data with socio-economic analysis. Field data collected between August and December 2023 from the Canarin and Scardovari Lagoons revealed seasonal fluctuations in crab abundance, with a peak in captures during the warmer months. The predatory behaviour of *C. sapidus* has led to a sharp decline in clam production, reaching near-zero levels in early 2024. Statistical analysis confirmed a strong correlation between the increase of the invasive crab population and the decrease in clam yields. This study also explores potential management strategies, including the economic valorisation of *C. sapidus* as a commercial resource, turning an ecological challenge into an opportunity. These findings highlight the urgent need for targeted management interventions to mitigate the impact of this invasive species on local fisheries and ecosystems.

## 1. Introduction

Biological invasions are one of the main threats for marine and transitional ecosystems. Invasive alien species (hereafter: IAS) are constantly increasing in the Mediterranean basin (e.g., Galil et al., 2008, 2017, 2018; Ulman et al., 2017; Servello et al., 2019; Katsanevakis et al., 2020; Zenetos et al., 2022; Ragkousis et al., 2023), and often generate significant ecological and economic impacts. The invasiveness of some marine alien species can compromise the functions and the structure of ecosystems and, at the same time, favor the onset of altered aquatic food webs and increase the biodiversity loss (Streftaris et al., 2005; Hulme, 2007; Vilà et al., 2011; Boudouresque & Verlaque, 2012; Seebens et al., 2017; Azzurro et al., 2019; Tsirintanis et al., 2022). Moreover, IAS impair the provision of ecosystem services resulting in economic losses for coastal human activities (e.g., fishery, tourism, agriculture, and breeding) (Cinar et al., 2014; Rai & Singh, 2020; Perez et al., 2022; Tsirintanis et al., 2022). Estuaries and lagoons are among the environments most affected by biological invasions as they are characterized by the presence of low hydrodynamism, anthropogenic pressure, pollution, and eutrophication (Occhipinti-Ambrogi & Savini, 2003; Sgro et al., 2005; Occhipinti-Ambrogi et al., 2011; Sfriso et al., 2012; Boudouresque et al., 2012; Marchini et al., 2015; Servello et al., 2019, Lacoste et al., 2023; Zamora-Marin et al., 2023).

Timely reporting of IAS is essential to promote effective forms of management in recipient areas; at the same time the implementation of proper management and mitigation policies are of fundamental importance to limit the geographic expansion of invasive species and to prevent the occurrence of direct impacts on human activities (Epstein & Smale, 2017; Giakoumi et al., 2019; Castro et al., 2021).

For example, proper capture and eradication programmes, such as the use of traps and selective fishing, can be considered key methods of invasive species management, as in the case of *Carcinus means* in South Africa (Mabin et al., 2020). Several studies have analyzed the interaction between IAS and human fishing activities evidencing that bioinvaders can exert a wide spectrum of negative impacts as e.g., decrease of landing, direct damage to catches and fishing gears, reduction of seafood quality, disease transmission, fouling, degradation of habitats (Latini & Petrere, 2004; Katsanevakis et al., 2009, 2018; Sala et al., 2011; Stebbing et al., 2012; Vergés et al., 2014; Fernandes et al., 2016; Corrales et al., 2017; Galil et al., 2021; Galván et al., 2022)

Comparable negative interactions can be attributed to the invasive crab *Callinectes sapidus*, whose Mediterranean populations are constantly expanding in recent years. The implementation of innovative management actions is crucial when dealing with invasive species that represent a threat to fisheries as well as a potential economic resource as a shell- or finfish product as in the case of blue crab (Mancinelli et al., 2017; De Carvalho-Souza et al., 2024).This Portunidae is native to the Western American coast and is considered a voracious and aggressive predator that lives in both marine and transitional environments (Galil et al., 2002; Mancinelli et al., 2017). The worrying expansion of this crab along the Italian and Mediterranean coasts has been favored by its biological characteristics such as: rapid growth rates, early maturity, opportunistic diet, large production of eggs, long-range larval dispersion, excellent swimming capacity and generalist habitat use (Tiralongo et al., 2021a; Clavero et al., 2022). These biological traits are typical of several species that generate a direct impact on biodiversity and fishing activities. Also, the invasions of the Snow Crab *Chinoecetes opilio* and the Red King Crab *Paralithodes camtschaticus*, that have threatened ecological balances and benthic food webs of Barrens Sea, showed a complex story which highlighted the presence of limitations of political, economic, and ecological management (Kaiser et al., 2018). However, despite this, over time these invasive crabs have become an important fishing resource for local fisheries (Agnalt et al., 2011; Falk-Petersen et al., 2011; Hansen, 2016; Sundet e Hoel, 2016; Lorentzen et al., 2018; Kourantidou & Kaiser, 2021).

In fact, the impact of IAS on biodiversity and human activities is the result of several different effects that can be both negative and positive (Katsanevakis et al., 2014; De Carvalho-Souza et al., 2024). In the context of edible species of high commercial interest, positive effects are associated with food provision as indicated by Galil (2007) for the case of some prawns belonging to the family Penaidae, considered IAS in the Mediterranean Sea. In recent years in fact, along the coasts of the eastern and central Mediterranean Sea, abundance, prizing and trawling catches of *Metapenaeus monoceros*, *Marsupenaeus japonicus*, *Penaeus aztecus*, and *P. semisulcatus* increased to become an economic and commercial resource of great importance (Galil, 2008; Spinelli et al., 2024). This form of approach has been applied to Lionfish, and this strategy boosted the demand, marketing and culinary potential of this IAS while speeding up the removal of the species (Green & Grosholz, 2021; Bogdanoff et al., 2021; Smith et al., 2023).

A concrete and functional approach is the market approach that involves the creation, promotion, and expansion of a market chain, useful to facilitate the containment of IAS. However, this approach must be supported by selective and effective fishing practices that are useful in significantly removing invasive species from the invaded environment (Hanley & Roberts; 2018; Kleitou et al., 2021; Smith et al., 2023). The implementation of an effective reform of non-indigenous species fisheries has the potential to facilitate the control and management of these fisheries, protect natural capital, stimulate the creation of new markets, and simultaneously ensure the restoration of ecosystems (Kleitou et al., 2021). These strategies can contribute to several goals related to the 2030 Agenda for Sustainable Development (e.g. Goal 14) and promoted by the Water Framework Directive (WFD Descriptor 2 - non-indigenous species).

The Italian Po Delta and its associated waters are subject to a multitude of anthropogenic activities, with the deltaic lagoons having a long history of hosting diverse fishing and farming practices (Cencini, 1998; Sorokin et al., 1999; Simeoni & Corbau, 2009; Bellafiore et al., 2019; Gaglio et al., 2019; Tamburini et al., 2019; Lanzoni et al., 2021).

The proliferation of mollusc farming facilities, the impact of anthropogenic stressors and the existence of numerous transport vectors have been identified as potential drivers of the recent settlement of allochthonous species in the Po Delta lagoon system (Mistri et al., 2004; Sgro et al., 2005; Sfriso et al., 2012; Cuesta et al., 2014; Munari et al., 2016; Di Blasio et al., 2023). In last years, the population of *C. sapidus* has been subject to a significant increase in the delta, resulting in adverse effects on shellfish farming and local natural ecosystems (Azzurro et al., 2024).

The aim of this study is to evaluate, through a multidisciplinary approach, the impact of *C. sapidus* on the economic activities of clam farming in the Po Delta, providing ecological and biological data on the population present there, discussing the socio-economic aspects, and suggesting possible solutions to the issue.

## 2. Material and methods

### 2.1 Study area

The study was carried out in two lagoons of the Po Delta (NE Italy): Canarin (centered at 44.92266N, 12.49370E) and Scardovari (centered at 44.85906N, 12.41786E). In general, these transitional water bodies are characterized by shallow-water environments with various levels of connection with the sea and with the river branches. Their hydro-morphological features, directly influenced by the outflows of the Po River, are highly dynamic and change rapidly due to biotic and abiotic forcings (Ceccherelli et al., 1994; Maicu et al., 2018; Franzoi et al., 2023). In addition, these lagoons are affected by several forms of anthropic pressure that over time have changed their natural ecological characteristics (Franzoi et al., 2023). For example, their basin hosts several Clam and Oyster farms, especially Pacific oyster (*Magallana gigas*) and Philippine clams (*Ruditapes philippinarum*), representing an important economic resource at regional and national scale (Turolla, 2008; Donati & Fabbro, 2010; Verza & Cattozzo, 2015; Bordignon et al., 2020).

Specifically, the Scardovari Lagoon extends over an area of about 32 km$^2$ and its waters have an average depth of 1.5 m (Mistri et al., 2018). This lagoon is connected with the sea through two inlets located in the North-East and in the South-West of the basin. The bottom substrates are mainly represented by mud and the tidal excursions are rather limited and reach a maximum of 1 m in the summer season (Munari et al., 2013; Bordignon et al., 2020). Freshwater inputs are represented by a small channel connecting to the Po di Gnocca and two pumping stations located in the NW bank of the lagoon (Maicu et al., 2018). Based on their different hydrological, hydrodynamic and sedimentological features the lagoon basin can be divided into two zones: the southern area, most influenced by the marine tides and the Po freshwater, and the northern area, where sea water inputs are minimal (Andreoli et al., 1994). For these reasons, the more confined northern areas show greater depths than the areas located to the south. In addition, the northern area is characterized by a higher nutrient input related to agricultural runoff, while the southern area, which is more influenced by exchanges with the sea, hosts numerous shellfish farms (Bordignon et al., 2020). Due to subsidence phenomena, which has affected the southern areas of the Po Delta, the Scardovari Lagoon has undergone a profound morphological change in the last century: the erosion of salt marshes and the filling of some connecting canals have led to a decrease in the lagoon surface area (more than 1.5 km) while the inner area of the lagoon has undergone a sharp deepening (Corbau et al., 2022).

The Canarin Lagoon is located in the southern area of the Delta, between the Pila and Tolle deltaic branches and is connected to the Adriatic Sea by a shallow mouth characterized by a width of about 200 m and a maximum depth of 2.5 m; this basin covers an area of about 6.4 km$^2$ and its average depth is of 0.9 m. This lagoon receives fresh water through three different inlets: from the Basson Lagoon with an opening located on the north bank of the basin, to the south through a small canal connected to the Po di Tolle, and to the west where a small amount of fresh water is pumped (Verza & Catozzo, 2015; Maicu et al., 2018). The basin is devoid of sandbanks and reed beds, and its body of water is mainly exploited for fishing activities and clam farming (Verza & Cattozzo, 2015). Moreover, this lagoon underwent numerous anthropogenic interventions useful for modifying the hydrodynamic characteristics of the basin; for example, to decrease the intake of fresh water, the northern section of the inlet was modified through the construction of a floodplain and a protective embankment (Verga & Catozzo, 2015).

*2.2 Sampling procedures*

The samplings of individual of blue crabs were conducted by researchers thanks to the support of local fishers using two fishing gears: 1) crab traps: cuboid-shaped and made of metal (60×60×40 cm), baited with the Thinlip grey mullet (Chelon ramada), and left in the lagoons for about 24 hours, at a depth of 0.4-1 m. Crab traps were controlled every day in the early morning, emptied and baited again. With this fishing gear, crabs were collected in the period extending from August 9$^{th}$ 2023 to December 22$^{nd}$ 2023; 2) ostreghero: it can be considered a small bottom trawl net with a rectangular mouth (140×50 cm) formed by a steel frame followed by a net and a small bag. On the sides of the metal mouth, there are two metal sleds, that make direct contact with the bottom. These sleds are connected by an iron chain longer than the distance between them, which helps to stir the bottom more and increase the catches of blue crabs. The net is dragged at a low speed, about one and a half knots, from the bow of the boat as it moves backward, at a depth of 0.5-1 m.

The crabs, once collected from the traps or the ostreghero's bag, were counted and weighed directly on board using a dynamometer. Due to the large quantity of crabs in some hauls, most of the samples taken with the ostreghero were estimated only by biomass in kilograms.

A total of 384 individuals were randomly selected during four sampling days widely distributed between August and November 2023 in the Canarin Lagoon. These specimens were used to collect measurements and biological parameters and were sampled using the "ostreghero net". The ostreghero net, which, unlike traps, eliminates the problem related to the poor selectivity of the latter. Smaller sizes, in fact, could pass through the mesh of the traps.

*2.3 Post-capture procedures*

The sex of crabs was easily and rapidly determined by observing external morphological characteristics, such as the shape of the abdomen, typically T-shaped in male specimens vs. a wider and rounder abdomen in females. These observations allowed us to stablish the sex-ratio (M:F) of studied population. Carapace width (CW) and carapace length (CL) were measured in millimeters (mm) using a digital caliper. The total fresh weight (W) of the crabs was determined using a precision scale, with values expressed in grams. The maturation stages of females were visually assessed based on the colour of eggs: S1 (bright yellow eggs, early maturation), S2 (maturing eggs, dark yellow-orange), and S3 (mature eggs, close to release, brownish-black). The weight of the egg masses (eggs plus abdomen) in grams was correlated with the size of females (W-Eggs weight) to determine any relationships between the females' weight and egg production. CW-W relationships were constructed for both sexes and combined. The relationship between weight (W) and carapace width (CW) was expressed using the allometric formula: $W = aCW^b$; where a and b are coefficients determined through linear regression on the log-transformed data, and better represented as a regression curve with non-log-transformed data. The population structure was analysed by dividing the crabs into size classes based on carapace width (CW). The size class distribution was determined for both sexes separately and combined to obtain a comprehensive view of the population.

*2.4 Data collection on clam yields and blue crab management*

The data on clam yields and the management of Atlantic blue crabs were obtained directly from the Scardovari Cooperative, located in the Po Delta region. The cooperative provided access to their internal records, which included detailed information on the monthly harvest volumes for claims and sales and disposal volumes of *C. sapidus*. These records were compiled and analysed to assess the economic impact of blue crab on the clam farming industry, as well as to evaluate the management strategies for its containment and possible valorisation. The quantities of crabs sold and disposed of, as well as those of clams in the various periods examined, were expressed in tons. Data on crabs were available starting from July 2023 and are updated to June 2024, for a whole period of 12 months. Data on monthly clams production were available starting from the year 2010 to date (June 2024).

*2.5 Data analysis*

For each fishing gear (ostreghero and crab traps), catches for unit efforts (CPUE) have been calculated. In the case of ostreghero, being comparable for most aspects to a small trawl net and considering the large abundance of crabs in the area, C.P.U.E. were estimated in kg/min (Tiralongo et al., 2021b). Catch per unit efforts for crab traps were instead measured in kg/fishing trap in a period of 24 hours. For this latter fishing gear, we also provided data on the number of crabs caught per fishing trap ($n_{crabs}$/fishing trap).

ANOVA was performed to establish if there were significant statistical differences in: i) size differences (CW) between males and females in different period of sampling; ii) abundance (number of individuals) and biomass (kg) in crabs catches per crab traps during different periods; iii) abundance in crab catches per minute with ostreghero net during different periods; iv) biomass (tonnes) of disposed, sold and total amount of crabs from the "Scardovari Cooperative" from July 2023 to June 2024.

Yields in clams production (tonnes) from the 2010 to 2024 (data updated to June 2024) were analysed through a Kruskal-Wallis Test for their non-parametric distribution.

A paired t-test was used to investigate significant difference between disposed and sold crabs during different periods, and to verify the existence of statistically significant difference in CW-W relationships between males and females of blue crabs. Chi-square test was used to evaluate the existence of statistically significant difference in the distribution between sexes, males vs. females.

## 3. Results and discussion

### 3.1 Yield in sampling periods (CPUE) of Callinectes sapidus in the Canarin Lagoon

The median number of crabs shows significant variations between periods, both for crab traps and "ostreghero net". With regard to crab traps, the median number of crabs (n crabs/crab trap) showed significant variations between periods, with high median values recorded in the period from August to October (Figure 1, Table 1). From the 1st of November, mean abundance of crabs per crab traps tend to significantly decrease. The 4th of October showed a significant peak in both biomass (kg crabs/crab trap) and number of crabs. For most other periods, the biomass remains low, indicating that captures mainly consisted of small-sized individuals. For this fishing gear, CPUE of crabs were expressed as crab kg/trap in 24 hr. Statistical analysis (one-way ANOVA) showed a clear difference (df=19, F=11.34, p<0.01) in yields depending on the period, with yields ranging from a mean value of 0.28±0.24 (in August) to 4.55±1.40 kg/trap (in October). As regard CPUE of $n_{crab}$/trap, ANOVA showed also in this case a clear difference in yields depending on the period (df=19, F=10.6, p<0.01), with yields ranging from a mean value of 2.8±2.14 to 29.7±8.72 $n_{crabs}$/trap (both results obtained in August).

With regard to "ostreghero net" the sampling period extended from August 9th to November 12nd. Also for this fishing gear, the crab biomass (kg crabs/min.) showed significant differences between sampling periods (Figure 2, Table 2). On the 16th of August, it was recorded a significant event of crab capture, and, after the other peaks observed in late August and mid-September, there was a significant decline in crab biomass capture until the end of sampling period in November. For this fishing gear, CPUE of crabs were expressed as crab kg/min. Statistical analysis (one-way ANOVA) showed a clear difference (df=12, F=17.18, p<0.01) in yields depending on the period, with yields ranging from a mean value of 0.17±0.13 (in November) to 9.60±4.37 kg/min (in August).

### 3.2 Impact of C. sapidus in clams production and data on disposal and sales of the crab

A sharp decline in local clam production during the analysed period (2010-2024) was observed at the end of 2023, and was clearly visible in 2024. This decline coincides with the proliferation of *C. sapidus*. Statistical analysis (Kruskal-Wallis test) showed a clear difference in the clams production over the time (df=14, H=76.407, p<0.01), from the year 2010 to 2024. A sharp decline is observable from the year 2018, with an increasingly marked reduction in conjunction with the increase in invasive crabs *C. sapidus*. Historic lows in clam production were observed from November 2023, with zero yield in the month of March and April 2024 (Figure 3).

Available data (July 2023-June 2024) on sales and disposals of blue crabs from lagoons of the Po Delta area showed a peak in disposal in August 2023, followed by a significant drop in subsequent months (Figure 4). On the other hand, sales show a more stable trend with minor fluctuations. The total volume (biomass) of blue crabs "production" (disposal plus sales) also demonstrates variations, with a marked peak in August 2023. Paired *t*-test showed that there was not a statistically significant difference between the biomass of disposed vs. sold crabs (df=11, *t*=1.323, p=0.213) in the period from July 2023 to June 2024. However, trend lines showed an opposite situation: upward for sold crabs and downward for disposed crabs. The month of August proved to be the most productive in terms of total tonnes of crabs caught; although similar results could also have been obtained in July, if crab fishing had not started systematically only from the second half of the month.

### 3.3 Allometric relationships and demographic structure for both sexes

For males, the allometric relationship between carapace width (CW) and weight (W) was expressed as W = 0.00027 × $CW^{2.72}$, with a $R^2$ value of 0.948, indicating a strong correlation between CW and W. For females, the relationship was expressed as: W = 0.00114 × $CW^{2.38}$ with an $R^2$ value of 0.891, also showing a strong correlation, tough slightly less than that observed in males (Figure 5). When considering both sexes combined, the relationship was: W = 0.00096 × $CW^{2.43}$ with a $R^2$ value of 0.943, reflecting a strong overall correlation.

The histogram of the combined population shows a bimodal distribution with two prominent peaks around 90-100 mm and 130-140 mm CW. The histogram for the female population also displays a bimodal distribution, with peaks around 80-100 mm and 130-150 mm CW. The histogram for the male population shows a unimodal distribution with a peak around 90-110 mm CW. With the exception of the last sampling day performed on November 26th, 2023, one-way ANOVA showed a clear statistical difference in the size between males and females, with the latter reaching larger sizes (df=375, F=62.05, p<0.01) (Figure 6, Table 3).

*3.4 Relationship eggs' mass vs. female weight, maturation stages, sex-ratio, and size at first maturity*

Results showed a positive correlation between female size (in this case weight) and eggs biomass, following the equation: EggsW = 0.24×W+3.39 (Figure 7). The Pearson's correlation coefficient was 0.79 with a significant p-value (p<0.01), indicating a statistically positive correlation between eggs and females' biomasses. Female dominance was recorded during both samplings performed in August, 1st and 19th, with a sex-ratio of 0.84 (p=0.28) and 0.67 (p=0.04), respectively. On the other hand, males were dominant in September and November, with a sex-ratio of 2.05 (p=0.006) and 1.38 (p=0.49), respectively. Ultimately, females were statistically more abundant on August 19th (p<0.05), while males were more abundant on September 1st (p<0.05). The analysis of the maturation stages distribution of *C. sapidus* over time revealed significant variations in the proportions of different developmental stages (S1, S2, and S3) in the collected samples. During the first sampling period on August 1st, 2023, the distribution of maturation stages showed that 25% of the samples were in stage 1 (S1), 50% in stage 2 (S2), and 25% in stage 3 (S3). This distribution indicates a predominance of the intermediate stage (S2), suggesting that a significant portion of the population was in an advanced stage of development. In the second sampling period on August 19th, 2023, the data showed a shift in proportions: 48.6% of the samples were in stage 1, 37.1% in stage 2, and 14.3% in stage 3. Compared to the previous period, there was a relative increase in stage 1 and a decrease in stage 3, suggesting the presence of a new cohort of younger individuals or regression in some individuals. Finally, the sampling on September 1st, 2023 revealed a distribution with 66.7% of the samples in stage 1 and 33.3% in stage 2, with no samples in stage 3. This shift towards a higher proportion of individuals in stage 1 and the absence of stage 3 individuals indicates a predominance of individuals in the early stages of maturation, potentially due to recent recruitment of young individuals.

*3.5 Discussion*

Seasonal variability is observed with abundance peaks in specific periods (August-October), suggesting potential environmental and/or behavioural influences that determine the availability and catchability of crabs with traps. Towards the later periods of sampling (November and December), although there is a slight decrease in biomass, it was observed the presence of many small-sized crabs. This could imply successful reproduction or recruitment of juvenile crabs into the population. A similar pattern was recorded for "ostreghero net" catches, in which crab biomass was higher in warmer months. This could be explained by a combination of behavioural and environmental factors. Concerning the latter ones, salinity, temperature and tidal action might influence crab movement and availability in the sampling areas. Moreover, the fact that crabs were never targeted and harvested massively until July 2023, certainly determined the large quantities caught in August. Compared to traps, the ostreghero net is significantly more effective in capturing a larger quantity of blue crabs in a shorter amount of time. It also captures all age classes, from very small juveniles to sexually mature adults of both sexes. In contrast, traps show a clear selectivity towards larger individuals. In conclusion, for blue crab removal, the ostreghero proves to be a more effective tool as it directly targets various cohorts within the population. However, its extensive and prolonged use could potentially have negative impacts on the habitat and native species.

The presence of *C. sapidus* in lagoons of the Po Delta area significantly reduced clam's production through direct predation. This was particularly evident starting from the end of 2023 and reached a production equal to zero in some months of the 2024, exactly March and April. Clams are an important economic resource in many Italian regions and the decline in their production represents a significant economic loss for fishermen and the entire supply chain (Donati & Fabbro, 2010; Martini et al., 2023). The sharp increase in crab's disposal during August 2023 suggests a critical event, demonstrating the rapid increase in *C. sapidus* abundance.

The subsequent decrease in disposal could indicate either a temporary reduction in blue crabs' abundance due to the fishing efforts or a shift in management strategies. On the other hand, sales of *C. sapidus* remain relatively stable with minor fluctuations, reflecting a consistent market demand for the species. The slight upward trend in sales towards the end of the period may indicate an adaptation by local markets to capitalize on the availability of blue crabs, partially turning a management challenge into an economic opportunity. Indeed, as has been observed in some lagoons in northern Egypt (Mehanna et al., 2019), an appropriate management of the *C. sapidus* fishery has the potential to transform this species into an economically valuable resource while simultaneously ensuring its demographic control, thereby favouring the re-establishment of shellfish farming and fishing activity in the Po Delta.

The total volume of blue crabs' yield showed a marked peak in August 2023, aligning with the disposal spike. This total volume decrease and later stabilization reflect the dynamic interplay between population control efforts, market demand, and environmental factors. In this context, employing integrated pest management strategies, including targeted fishing and commercial exploitation can incentive continuous and consistent removal of blue crabs, turning an ecological challenge into a market opportunity.

The allometric relationships between carapace width (CW) and weight (W) in *C. sapidus* demonstrate strong correlations, as indicated by the high $R^2$ in all cases. The different exponents (coefficients b) in the equation for males (2.72) and females (2.38) suggested sexual dimorphism in growth patterns, with males exhibiting a slightly steeper increase in weight relative to carapace width compared to females. These findings are crucial for understanding the growth dynamics and population structure of the invasive blue crab in the Po Delta, aiding in the management and control of this species.

Moreover, the strong correlation between CW and W supports the use of these morphometric parameters in ecological and fisheries research for monitoring the health and dynamics of crab populations.

The sex-ratio of *C. sapidus* in the investigated areas exhibited significant temporal variation across the different sampling periods, indicating fluctuations in population structure throughout the year. During August, there was a noticeable female dominance, in contrast, the sex-ratio shifted towards male dominance in the later months, although the difference was statistically significant only on August 19th (females' dominance) and on September 1st (males' dominance). This could indicate a temporal shift in the population structure, possibly due to behavioural or environmental factors affecting the distribution and visibility of the sexes. Moreover, the dominance of female individuals might also be attributed to the practice of releasing less commercially valuable females by fishermen, who prefer retaining larger male individuals.

This human alteration is likely the main factor that led to a skewed sex ratio in favour of females during August. In fact, starting from July, this practice was discouraged, and females were also retained for sale and, especially, disposal. This caused the sex ratio to stabilize in favour of males over time, as is clearly showed in the sex-ratio results in the month of September. The equation obtained from the W-EggsWeight relationships in female suggests that for every gram increase in the female's weight, the weight of the egg mass increases by 0.24 g. The positive slope of the regression line indicates a direct relationship between female body weight and egg mass weight.

This is consistent with the general understanding that larger individuals have more resources to allocate to reproduction, resulting in higher fecundity. Moreover, despite the overall positive trend, there is noticeable variability in the egg mass weights for females of similar body weights. This variability could be attributed to several factors, including differences in individual health, age, environmental conditions and genetic factors. Further studies could explore these variables to better understand factors influencing reproductive output. In all cases, the specific removal of large females results to be of greatest relevance, with potential long-term impacts on the *C. sapidus* population in the area.

Size distribution of female specimens with an evident bimodal distribution suggests that females are present in multiple size classes, with a significant number of large individuals. The presence of larger females aligns with the earlier observation that females with greater body size produce larger egg masses, contributing significantly to the reproductive potential of the population. The absence of a second peak in size distribution of male specimens, as observed in the females, suggests that the male population may have a more uniform size distribution or that larger males are less prevalent because targeted in the pre-invasion period.

This could be due to natural mortality, predation, or fishing pressures that selectively remove larger males from the population. Overall, the bimodal distribution for combined sexes suggests the presence of at least two distinct size classes within the population, possibly corresponding to different age groups and cohorts. Concerning sizes of individuals sampled, data showed a general trend where female consistently have larger median CW compared to males in August, but by September, the median CW of males increases significantly, surpassing that of females. In all cases, there is a noticeable fluctuation in the CW of both males and females across the sampling periods.

Females show a relatively stable size distribution with slightly higher medians in August and September, whereas males exhibit a significant increase in CW in November. The lower CW observed in males during the first three sampling periods between August and September might indicate a relatively long period of exploitation by the latter, as commented above. Moreover, the larger CW in females during the reproductive season (August) suggests that the reproductive females tend to be larger, which is consistent with the need for greater energy reserves for egg production.

Furthermore, this is likely due, at least in part, to the fact that during the pre-invasion period, female specimens carrying eggs were routinely released back into the sea and not retained by fishermen. This practice may have contributed to the current population dynamics, allowing a higher reproductive success and contributing to the rapid expansion of the species. The variations in the proportions of maturation stages across different periods suggest temporal dynamics in the *C. sapidus* population in the Canarin Lagoon. The high number of individuals in stages 2 and 3 at the beginning of August might reflect favourable environmental conditions for advanced maturation.

The invasion of *C. sapidus* depicts significant ecological and economic challenges; however, it also offers an opportunity to develop innovative solutions for waste utilization. Converting blue crab waste into valuable resources, such as chitin and chitosan, can address environmental concerns, promote sustainability, and create economic benefits. Nevertheless, the immediate priority must be the removal of this species from the Po Delta lagoons to protect the crucial clam farming industry. Clam farming is one of the most important economic activities in the Po Delta.

The invasion of the blue crab poses a direct threat to this industry, as the predatory crab decimated clam populations, leading to significant economic losses. Therefore, it is essential to implement removal and control strategies to limit the spread of the blue crab in these critical areas. In addition to removal efforts, utilizing blue crab waste for the production of chitin and chitosan represents an opportunity to mitigate the environmental impacts of the invasion.

These biopolymers have numerous industrial, agricultural, and biomedical applications that can stimulate the local economy through the creation of new industries and jobs. Finally, investment in research and development will be crucial for maximizing the potential of blue crab waste utilization and ensuring the long-term success of these initiatives. Collaborations between research institutions, industries, and local communities will be essential for developing sustainable technologies and effective strategies for managing invasive species and protecting local economic activities, such as clam farming in the Po Delta.

*3.6 Conclusions*

The invasion of the blue crab *Callinectes sapidus* in the Po Delta lagoons presents significant ecological and economic challenges. The abundance of the crab exhibits seasonal variability, with capture peaks observed between August and October 2023, suggesting environmental and behavioural influences that determine its availability. The presence of many small-sized crabs in November and December 2023 indicates successful recruitment of juveniles into the population. This phenomenon, coupled with the reduction in clam production, which reached near-zero levels in some months of 2024, highlights the direct and heavy impact of *C. sapidus* on local fishing activities and economy.

Catches using the "ostreghero" net show a similar pattern to those of crab traps, although the ostreghero net yields are considerably higher. The highest biomass of crabs was recorded in warmer months, likely influenced by environmental and fishing-related factors. Additionally, the critical increase in crab disposal in August 2023 suggests an exceptional event, probably related to the rapid increase in the species' abundance, followed by a subsequent decrease, potentially due to fishing efforts or shifts in management strategies. However, sales of *C. sapidus* have remained stable with minor fluctuations, suggesting an adaptation of the local market to the abundance of the crab, partially turning a management challenge into an economic opportunity.

Variations in sex ratios and the size distribution of sampled individuals suggest temporal dynamics in the *C. sapidus* population, with a female dominance in the summer months followed by a stabilization favouring males. This trend can be attributed to both behavioural and environmental factors as well as fishing practices, which have influenced the population structure. The removal of large females, in particular, is crucial for containing the *C. sapidus* population, finally protecting the clam farming industry in the Po Delta.

In conclusion, while *C. sapidus* poses a significant threat to the local ecosystem, it is imperative to implement effective and targeted fishing strategies aimed at significantly reducing the abundance of the blue crab. This approach is essential not only to mitigate the ecological impact but also to allow the recovery of clam production in the area, ensuring the sustainability of this vital economic resource.

**Acknowledgements**

The authors are truly grateful to the Scardovari fishermen for their precious collaboration during the field sampling and the field information provided by them as part of this research.

**References**

Agnalt, A. L., Pavlov, V., Jørstad, K. E., Farestveit, E., & Sundet, J. (2011). The snow crab, *Chionoecetes opilio* (Decapoda, Majoidea, Oregoniidae) in the Barents Sea. In the wrong place- In the Wrong Place - Alien Marine Crustaceans: Distribution, Biology and Impacts (pp. 283-300). Springer.

Andreoli, C., Tolomio, C., Scarabel, L., Moro, I., Bellato, S., Moretto, M., & Masiero, L. (1994). Phytoplankton and chemico-physical parameters of the Scardovari Lagoon (Po Delta, North Adriatic Sea) during 1991 and 1992. Plant Biosystem, 128(6), 1007-1027.

Anger, K., Queiroga, H., & Calado, R. (2015). Larval development and behaviour strategies in Brachyura. In Treatise on Zoology-Anatomy, Taxonomy, Biology. The Crustacea, Volume 9 Part C (2 vols) (pp. 317-374). Brill.

Azra, M. N., Aaqillah-Amr, M. A., Ikhwanuddin, M., Ma, H., Waiho, K., Ostrensky, A.,Tavares C. P. S., & Abol-Munafi, A. B. (2020). Effects of climate-induced water temperature changes on the life history of brachyuran crabs. Reviews in Aquaculture, 12(2), 1211-1216.

Azzurro, E., Sbragaglia, V., Cerri, J., Bariche, M., Bolognini, L., Ben Souissi, J., Busoni, G., Coco, S., Chryssanthi, A., Fanelli, E., Ghanem, R., Garrabou, J., Gianni, F., Grati, F., Kolitari, J., Letterio, G., Lipej, L., Mazzoldi, C., Milone, N., … & Moschella, P. (2019). Climate change, biological invasions, and the shifting distribution of Mediterranean fishes: A large-scale survey based on local ecological knowledge. Global Change Biology, 25(8), 2779-2792.

Azzurro, E., Bonanomi, S., Chiappi, M., De Marco, R., Luna, G. M., Cella, M., ... & Strafella, P. (2024). Uncovering unmet demand and key insights for the invasive blue crab (*Callinectes sapidus*) market before and after the Italian outbreak: Implications for policymakers and industry stakeholders. Marine Policy, 167, 106295.

Bellafiore, D., Ferrarin, C., Braga, F., Zaggia, L., Maicu, F., Lorenzetti, G., ... & De Pascalis, F. (2019). Coastal mixing in multiple-mouth deltas: A case study in the Po delta, Italy. Estuarine, Coastal and Shelf Science, 226, 106254.

Bogdanoff, A. K., Shertzer, K. W., Layman, C. A., Chapman, J. K., Fruitema, M. L., Solomon, J., Sabattis, J., Green, S., & Morris, J. A. (2021). Optimum lionfish yield: a non-traditional management concept for invasive lionfish (*Pterois* spp.) fisheries. Biological Invasions, 23, 795-810.

Bordignon, F., Zomeño, C., Xiccato, G., Birolo, M., Pascual, A., & Trocino, A. (2020). Effect of emersion time on growth, mortality and quality of Pacific oysters (*Crassostrea gigas*, Thunberg 1973) reared in a suspended system in a lagoon in Northern Italy. Aquaculture, 528, 735481.

Boudouresque, C. F., & Verlaque, M. (2012). An overview of species introduction and invasion processes in marine and coastal lagoon habitats. CBM-Cahiers de Biologie Marine, 53(3), 309.

Castejón, D., Rotllant, G., Giménez, L., Torres, G., & Guerao, G. (2015). The effects of temperature and salinity on the survival, growth and duration of the larval development of the common spider crab *Maja brachydactyla* (Balss, 1922) (Brachyura: Majidae). Journal of Shellfish Research, 34(3), 1073-1083.

Castro, K. L., Battini, N., Giachetti, C. B., Trovant, B., Abelando, M., Basso, N. G., & Schwindt, E. (2021). Early detection of marine invasive species following the deployment of an artificial reef: Integrating tools to assist the decision-making process. Journal of Environmental Management, 297, 113333.

Ceccherelli, V. U., Ferrari, I., & Viaroli, P. (1994). Ecological research on the animal communities of the Po River Delta lagoons. Bollettino di Zoologia, 61(4), 425-436.

Cencini, C. (1998). Physical processes and human activities in the evolution of the Po delta, Italy. Journal of Coastal Research, 775-793.

Çinar, M. E., Arianoutsou, M., Zenetos, A., & Golani, D. (2014). Impacts of invasive alien marine species on ecosystem services and biodiversity: a pan-European review. Aquatic Invasions, 9(4), 391-423.

Clavero, M., Franch, N., Bernardo-Madrid, R., López, V., Abelló, P., Queral, J. M., & Mancinelli, G. (2022). Severe, rapid and widespread impacts of an Atlantic blue crab invasion. Marine Pollution Bulletin, 176, 113479.

Corbau, C., Zambello, E., Nardin, W., & Simeoni, U. (2022). Secular diachronic analysis of coastal marshes and lagoons evolution: Study case of the Po river delta (Italy). Estuarine, Coastal and Shelf Science, 268, 107781.

Corrales, X., Ofir, E., Coll, M., Goren, M., Edelist, D., Heymans, J. J., & Gal, G. (2017). Modeling the role and impact of alien species and fisheries on the Israeli marine continental shelf ecosystem. Journal of Marine Systems, 170, 88-102.

Costlow, J. D. (1967). The effect of salinity and temperature on survival and metamorphosis of megalops of the blue crab *Callinectes sapidus*. Helgoländer wissenschaftliche Meeresuntersuchungen, 15(1), 84-97.

De Carvalho-Souza, G. F., Kourantidou, M., Laiz, L., Nuñez, M. A., & González-Ortegón, E. (2024). How to deal with invasive species that have high economic value? Biological Conservation, 292: 110548.

Donati F., & Fabbro E. (2010). La molluschicoltura nelle lagune del Delta del Po veneto. Aspetti socio - economici. Relazione tecnica Consorzio di Bonifica Delta del Po.

Epstein, G., & Smale, D. A. (2017). Undaria pinnatifida: A case study to highlight challenges in marine invasion ecology and management. Ecology and Evolution, 7(20), 8624-8642.

Falk-Petersen, J., Renaud, P., & Anisimova, N. (2011). Establishment and ecosystem effects of the alien invasive red king crab (*Paralithodes camtschaticus*) in the Barents Sea–a review. ICES Journal of Marine Science, 68(3), 479-488.

Fernandes, J. A., Santos, L., Vance, T., Fileman, T., Smith, D., Bishop, J. D., Viard, F., Queiós, A. M., Merino, G.,

Buisman, E., & Austen, M. C. (2016). Costs and benefits to European shipping of ballast-water and hull-fouling treatment: Impacts of native and non-indigenous species. Marine Policy, 64, 148-155.

Fischer, S., Thatje, S., & Brey, T. (2009). Early egg traits in *Cancer setosus* (Decapoda, Brachyura): effects of temperature and female size. Marine Ecology Progress Series, 377, 193-202.

Franzoi, P., Facca, C., Redolfi Bristol, S., Boschiero, M., Matteo, Z., & Scapin, L. (2023). Application of the Habitat Fish Biological Index (HFBI) for the assessment of the ecological status of Po Delta lagoons (Italy). Italian Journal of Freshwater Ichthyology, 9, 91-106.

Gaglio, M., Lanzoni, M., Nobili, G., Viviani, D., Castaldelli, G., & Fano, E. A. (2019). Ecosystem services approach for sustainable governance in a brackish water lagoon used for aquaculture. Journal of Environmental Planning and Management, 62(9), 1501-1524.

Galil, B., Froglia, C., Noël, P. (2002). CIESM Atlas of Exotic Species in the Mediterranean. Vol 2. Crustaceans: decapods and stomatopods. Briand F. (ed), CIESM Publishers, Monaco, 192 pp.

Galil, B. S. (2008). Alien species in the Mediterranean Sea—which, when, where, why?. Challenges to marine ecosystems (pp. 105-116). Springer, Dordrecht.

Galil, B., Marchini, A., Occhipinti-Ambrogi, A., & Ojaveer, H. (2017). The enlargement of the Suez Canal—Erythraean introductions and management challenges. Management of Biological Invasions, 8(2), 141-152.

Galil, B. S., Marchini, A., & Occhipinti-Ambrogi, A. (2018). East is east and West is west? Management of marine bioinvasions in the Mediterranean Sea. Estuarine, Coastal and Shelf Science, 201, 7-16.

Galil, B. S., Mienis, H. K., Hoffman, R., & Goren, M. (2021). Non-indigenous species along the Israeli Mediterranean coast: tally, policy, outlook. Hydrobiologia, 848, 2011-2029.

Galván, D. E., Bovcon, N. D., Cochia, P. D., González, R. A., Lattuca, M. E., Reinaldo, M. O., Rincón-Díaz, M. P., Romero, M. A., Vanella, F. A., Venerus, L. A., & Svendsen, G. M. (2022). Changes in the specific and biogeographic composition of coastal fish assemblages in Patagonia, driven by climate change, fishing, and invasion by alien species. Global change in Atlantic coastal Patagonian ecosystems: A journey through time (pp. 205-231). Springer.

Giakoumi, S., Katsanevakis, S., Albano, P. G., Azzurro, E., Cardoso, A. C., Cebrian, E., Deidun, A., Edelist, D., Francour, P., Jimenez, C., Macic, V., Occhipinti-Ambrogi, A., Rilov, G., & Sghaier, Y. R. (2019). Management priorities for marine invasive species. Science of the Total Environment, 688, 976-982.

Glamuzina, L., Conides, A., Mancinelli, G., & Glamuzina, B. (2021). A comparison of traditional and locally novel fishing gear for the exploitation of the invasive Atlantic blue crab in the Eastern Adriatic Sea. Journal of Marine Science and Engineering 9(9), 1019.

Green, S. J., & Grosholz, E. D. (2021). Functional eradication as a framework for invasive species control. Frontiers in Ecology and the Environment, 19(2), 98-107.

Gu, D. E., Ma, G. M., Zhu, Y. J., Xu, M., Luo, D., Li, Wei, H., Xidong, M., Luo, J. R., & Hu, Y. C. (2015). The impacts of invasive Nile tilapia (*Oreochromis niloticus*) on the fisheries in the main rivers of Guangdong Province, China. Biochemical Systematics and Ecology, 59, 1-7.

Hanley, N., & Roberts, M. (2019). The economic benefits of invasive species management. People and Nature, 1(2), 124-137.

Hansen, H. S. B. (2016). Three major challenges in managing non-native sedentary Barents Sea snow crab (*Chionoecetes opilio*). Marine Policy, 71, 38-43.

He, P., Chopin, F., Suuronen, P., Ferro, R.S.T. and Lansley, J. 2021. Classification and illustrated definition of fishing gears. FAO Fisheries and Aquaculture Technical Paper 672.

Hernández, J. E., Palazón-Fernández, J. L., Hernández, G., & Bolaños, J. (2012). The effect of temperature and salinity on the larval development of *Stenorhynchus seticornis* (Brachyura: Inachidae) reared in the laboratory. Journal of the Marine Biological Association of the United Kingdom, 92(3), 505-513.

Hines, A. H. (2007). Chapter 14: Ecology of juvenile and adult blue crabs. Biology of the Blue Crab, edited by Kenney, V. S. and E. Cronin, 575–665. College Park, MD. Maryland Sea Grant Program.

Hulme, P. E. (2007). Biological invasions in Europe: drivers, pressures, states, impacts and responses. Biodiversity under Threat, 25, 56-80.

Kaiser, B. A., Kourantidou, M., & Fernandez, L. (2018). A case for the commons: the snow crab in the Barents. Journal of Environmental Management, 210, 338-348.

Katsanevakis, S., Tsiamis, K., Ioannou, G., Michailidis, N., & Zenetos, A. (2009). Inventory of alien marine species of Cyprus (2009). Mediterranean Marine Science, 10(2), 109-134.

Katsanevakis, S., Coll, M., Piroddi, C., Steenbeek, J., Ben Rais Lasram, F., Zenetos, A., & Cardoso, A. C. (2014). Invading the Mediterranean Sea: biodiversity patterns shaped by human activities. Frontiers in Marine Science, 1, 32.

Katsanevakis, S., Rilov, G., & Edelist, D. (2018). Impacts of marine invasive alien species on European fisheries and aquaculture–plague or boon?. CIESM Monograph, 50, 125-132.

Katsanevakis, S., Poursanidis, D., Hoffman, R., Rizgalla, J., Rothman, S. B. S., Levitt-Barmats, Y. A., Hadjioannou, L., Trkov, D., Garmendia, J. M., Rizzo, M., Bartolo, A., Bariche, M., Tomas, F., Kleitou, P., Schembri, P. J., Kletou, D., Tiralongo, F., Pergent-Martini, C., Pergent, G., … & Zenetos, A. (2020). Unpublished Mediterranean records of marine alien and cryptogenic species. BioInvasions Records, 9(2), 165-182.

Kleitou, P., Crocetta, F., Giakoumi, S., Giovos, I., Hall-Spencer, J. M., Kalogirou, S., ... & Rees, S. (2021). Fishery reforms for the management of non-indigenous species. Journal of Environmental Management, 280, 111690.

Kourantidou, M., & Kaiser, B. A. (2021b). Allocation of research resources for commercially valuable invasions: Norway's red king crab fishery. Fisheries Research, 237, 105871.

Lacoste, É., Jones, A., Callier, M., Klein, J., Lagarde, F., & Derolez, V. (2023). A Review of Knowledge on the Impacts of Multiple Anthropogenic Pressures on the Soft-Bottom Benthic Ecosystem in Mediterranean Coastal Lagoons. Estuaries and Coasts, 1-18.

Lanzoni, M., Gaglio, M., Gavioli, A., Fano, E. A., & Castaldelli, G. (2021). Seasonal variation of functional traits in the fish community in a brackish lagoon of the Po River Delta (northern Italy). Water, 13(5), 679.

Lárez, M. B., Palazón-Fernández, J. L., & Bolaños, C. J. (2000). The effect of salinity and temperature on the larval development of *Mithrax caribbaeus* Rathbun, 1920 (Brachyura: Majidae) reared in the laboratory. Journal of Plankton Research, 22(10), 1855-1869.

Latini, A. O., & Petrere Jr, M. (2004). Reduction of a native fish fauna by alien species: an example from Brazilian freshwater tropical lakes. Fisheries Management and Ecology, 11(2), 71-79.

Lorentzen, G., Voldnes, G., Whitaker, R. D., Kvalvik, I., Vang, B., Gjerp Solstad, R., Thomassen, M. R., & Siikavuopio, S. I. (2018). Current status of the red king crab (*Paralithodes camtchaticus*) and snow crab (*Chionoecetes opilio*) industries in Norway. Reviews in Fisheries Science & Aquaculture, 26(1), 42-54.

Mabin, C. A., Wilson, J. R., Le Roux, J. J., Majiedt, P., & Robinson, T. B. (2020). The first management of a marine invader in Africa: The importance of trials prior to setting long-term management goals. Journal of Environmental Management, 261, 110213.

Maicu, F., De Pascalis, F., Ferrarin, C., & Umgiesser, G. (2018). Hydrodynamics of the Po River-Delta-Sea System. Journal of Geophysical Research: Oceans, 123(9), 6349-6372.

Mancinelli, G., Chainho, P., Cilenti, L., Falco, S., Kapiris, K., Katselis, G., Ribeiro, F. (2017). The Atlantic blue crab *Callinectes sapidus* in southern European coastal waters: Distribution, impact and prospective invasion management strategies. Marine Pollution Bulletin 119: 5–11.

Mancinelli, G., Carrozzo, L., Costantini, M. L., Rossi, L., Marini, G., & Pinna, M. (2013). Occurrence of the Atlantic blue crab *Callinectes sapidus* Rathbun, 1896 in two Mediterranean coastal habitats: Temporary visitor or permanent resident?. Estuarine, Coastal and Shelf Science, 135, 46-56.

Marchini, A., Ferrario, J., Sfriso, A., & Occhipinti-Ambrogi, A. (2015). Current status and trends of biological invasions in the Lagoon of Venice, a hotspot of marine NIS introductions in the Mediterranean Sea. Biological invasions, 17(10), 2943-2962.

Martelli, A., & Baron, P. J. (2021). Effects of temperature and salinity on the development and survival of the embryos and zoeae I from the southern surf crab *Ovalipes trimaculatus* (Brachyura: Portunidae). Anais da Academia Brasileira de Ciências, 93, e20190999.

Martin, C. W., Valentine, M. M., & Valentine, J. F. (2010). Competitive interactions between invasive Nile tilapia and native fish: the potential for altered trophic exchange and modification of food webs. PLoS ONE, 5(12), e14395.

Martini, A., Aguiari, L., Capoccioni, F., Martinoli, M., Napolitano, R., Pirlo G., Tonachella, N., & Pulcini, D. (2023). Is Manila Clam farming environmentally sustainable? A life cycle assessment (LCA) approach applied to an Italian *Ruditapes philippinarum* hatchery. Sustainability, 15(4), 3237.

Mehanna, S. F., G Desouky, M., & E Farouk, A. (2019). Population dynamics and fisheries characteristics of the Blue Crab *Callinectes sapidus* (Rathbun, 1896) as an invasive species in Bardawil Lagoon, Egypt. Egyptian Journal of Aquatic Biology and Fisheries, 23(2), 599-611.

Mistri, M., Borja, A., Aleffi, I. F., Lardicci, C., Tagliapietra, D., & Munari, C. (2018). Assessing the ecological status of Italian lagoons using a biomass-based index. Marine Pollution Bulletin, 126, 600-605.

Munari, C., Rossetti, E., & Mistri, M. (2013). Shell formation in cultivated bivalves cannot be part of carbon trading systems: a study case with *Mytilus galloprovincialis*. Marine Environmental Research, 92, 264-267.

Occhipinti-Ambrogi, A., & Savini, D. (2003). Biological invasions as a component of global change in stressed marine ecosystems. Marine Pollution Bulletin, 46(5), 542-551.

Occhipinti-Ambrogi, A., Marchini, A., Cantone, G., Castelli, A., Chimenz, C., Cormaci, M., Froglia, C., Gambi, M. C., Giaccone, G., Giangrande, A., Gravili, C., Mastrototaro, F., Mazziotti, C., Orsi-Relini, L., & Piraino, S. (2011). Alien species along the Italian coasts: an overview. Biological Invasions, 13(1), 215-237.

Papadopoulos, I., Newman, B. K., Schoeman, D. S., & Wooldridge, T. H. (2006). Influence of salinity and temperature on the larval development of the crown crab, *Hymenosoma orbiculare* (Crustacea: Brachyura: Hymenosomatidae). African Journal of Aquatic Science, 31(1), 43-52.

Paula, J., Mendes, R. N., Mwaluma, J., Raedig, C., & Emmerson, W. (2003). Combined effects of temperature and salinity on larval development of the mangrove crab *Parasesarma catenata* Ortman, 1897 (Brachyura: Sesarmidae). Western Indian Ocean Journal of Marine Science, 2(1), 57-63.

Pérez, G., Vila, M., & Gallardo, B. (2022). Potential impact of four invasive alien plants on the provision of ecosystem services in Europe under present and future climatic scenarios. Ecosystem Services, 56, 101459.

Ragkousis, M., Zenetos, A., Souissi, J. B., Hoffman, R., Ghanem, R., Taşkın, E., Muresan, M., Karpova, E., Slynko, E., Dagli, E., Fortic, A., Surugiu, V., Macic, V., Trkov, D., Rjiba-Bahri, W., Tsiamis, K., Ramos-Espia, A. A., Petovic, S., Ferrario, J., … & Katsanevakis, S. (2023). Unpublished Mediterranean and Black Sea records of marine alien, cryptogenic, and neonative species. BioInvasions Records, 12(2), 339-369.

Rai, P. K., & Singh, J. S. (2020). Invasive alien plant species: Their impact on environment, ecosystem services and human health. Ecological Indicators, 111, 106020.

Ravi, R., & Manisseri, M. K. (2012). Survival rate and development period of the larvae of *Portunus pelagicus* (Decapoda, Brachyura, Portunidae) in relation to temperature and salinity. Fisheries and Aquaculture Journal, vol. 2012, FAJ-49, 1-9.

Romero, M. C., Tapella, F., Stevens, B., & Loren Buck, C. (2010). Effects of reproductive stage and temperature on rates of oxygen consumption in *Paralithodes platypus* (Decapoda: Anomura). Journal of Crustacean Biology, 30(3), 393-400.

Seebens, H., Blackburn, T. M., Dyer, E. E., Genovesi, P., Hulme, P. E., Jeschke, J. M., Pagad, S., Pysek, P., Winter, M., Arianoutsou, M., Bacher, S., Blasius, B., Brundu, G., Capinha, C., Celesti-Grapow, L., Dawson, W., Dullinger, S., Fuentes, N., Jager, H., … & Essl, F. (2017). No saturation in the accumulation of alien species worldwide. Nature communications, 8(1), 14435.

Servello, G., Andaloro, F., Azzurro, E., Castriota, L., Catra, M., Chiarore, A., Crocetta, F., D'Alessandro, M., Denitto, F., Froglia, C., Gravili, C., Langer, M. R., Lo Brutto, S., Mastrototaro, F., Petrocelli, A., Pipitone, C., Piraino, S., Relini, G., Serio, D., … & Zenetos, A. (2019). Marine alien species in Italy: A contribution to the implementation of descriptor D2 of the marine strategy framework directive. Mediterranean Marine Science, 20(1), 1-48.

Sfriso, A., Wolf, M. A., Maistro, S., Sciuto, K., & Moro, I. (2012). Spreading and autoecology of the invasive species *Gracilaria vermiculophylla* (Gracilariales, Rhodophyta) in the lagoons of the north-western Adriatic Sea (Mediterranean Sea, Italy). Estuarine, Coastal and Shelf Science, 114, 192-198.

Sgro, L., Munari, C., Angonese, A., Basso, S., & Mistri, M. (2005). Functional responses and scope for growth of two non-indigenous bivalve species in the Sacca di Goro (northern Adriatic Sea, Italy). Italian Journal of Zoology, 72(3), 235-239.

Shuai, F., Li, J., & Lek, S. (2023). Nile tilapia (*Oreochromis niloticus*) invasion impacts trophic position and resource use of commercially harvested piscivorous fishes in a large subtropical river. Ecological Processes, 12(1), 1-17.

Simeoni, U., & Corbau, C. (2009). A review of the Delta Po evolution (Italy) related to climatic changes and human impacts. Geomorphology, 107(1-2), 64-71.

Smith, N., Burgess, K., Clements, K. R., Burgess, J. C., Lavoie, A., & Solomon, J. N. (2023). Serving conservation from reef to plate: Barriers and opportunities for invasive lionfish consumption in restaurants. Aquatic Conservation: Marine and Freshwater Ecosystems, 33(6), 566-578.

Sorokin, Y. I., Sorokin, P. Y., & Ravagnan, G. (1999). Analysis of lagoonal ecosystems in the Po River Delta associated with intensive aquaculture. Estuarine, Coastal and Shelf Science, 48(3), 325-341.

Spinelli, A., Baquero, P. S., & Tiralongo, F. (2024). Westward expansion of the brown shrimp *Penaeus aztecus* Ives 1891 (Decapoda: Penaeidae) in the Mediterranean Sea: a review on the Mediterranean distribution and first record from Spain. Natural History Sciences, 11(1), 65-70.

Stebbing, P. D., Pond, M. J., Peeler, E., Small, H. J., Greenwood, S. J., & Verner-Jeffreys, D. (2012). Limited prevalance of gaffkaemia (*Aerococcus viridans* var. *homari*) isolated from wild-caught European lobsters *Homarus gammarus* in England and Wales. Diseases of Aquatic Organisms, 100(2), 159-167.

Streftaris, N.; Zenetos, A., & Papathanassiou, E. (2005). Globalisation in marine ecosystems: the story of non-indigenous marine species across European seas. Oceanogry and Marine Biology: an Annual Review. 43: 419-453.

Sundet, J. H., & Hoel, A. H. (2016). The Norwegian management of an introduced species: the Arctic red king crab fishery. Marine Policy, 72, 278-284.

Tamburini, E., Fano, E. A., Castaldelli, G., & Turolla, E. (2019). Life cycle assessment of oyster farming in the Po Delta, Northern Italy. Resources, 8(4), 170.

Tiralongo, F., Villani, G., Arciprete, R., & Mancini, E. (2021a). Filling the gap on Italian records of an invasive species: first records of the Blue Crab, *Callinectes sapidus* Rathbun, 1896 (Decapoda: Brachyura: Portunidae), in Latium and Campania (Tyrrhenian Sea). Acta Adriatica, 62(1), 99-104.

Tiralongo, F., Mancini, E., Ventura, D., De Malerbe, S., DE MENDOZA, F. P., Sardone, M., ... & Minervini, R. (2021b). Commercial catches and discards composition in the central Tyrrhenian Sea: a multispecies quantitative and qualitative analysis from shallow and deep bottom trawling. Mediterranean Marine Science, 22(3), 521-531.

Tsirintanis, K., Azzurro, E., Crocetta, F., Dimiza, M., Froglia, C., Gerovasileiou, V., Langeneck, J., Mancinelli, G., Rosso, A., Stern, N., Triantaphyllou, M., Tsiamis, K., Turon, X., Veriaque, M., Zenetos, M., & Katsanevakis, S. (2022). Bioinvasion impacts on biodiversity, ecosystem services, and human health in the Mediterranean Sea. Aquatic Invasions, 17(3), 308-352.

Turolla, E. (2008). La venericoltura in Italia. Estado actual de cultivo y manejo de moluscos bivalvos y su proyección futura: factores que afectan su sustentabilidad en América Latina, edited by Lovatelli, A., Farías, A., Uriarte, I. Taller Técnico Regional de la FAO. 20–24 de agosto de 2007, Puerto Montt, Chile. FAO Actas de Pesca y Acuicultura. No. 12. Roma, FAO. pp. 177–188.

Ulman, A., Ferrario, J., Occhpinti-Ambrogi, A., Arvanitidis, C., Bandi, A., Bertolino, M., ... & Marchini, A. (2017). A massive update of non-indigenous species records in Mediterranean marinas. PeerJ, 5, e3954.

Valdes, L., Alvarez-Ossorio, M. T., & Gonzalez-Gurriaran, E. (1991). Influence of temperature on embryonic and larval development in *Necora puber* (Brachyura, Portunidae). Journal of the Marine Biological Association of the United Kingdom, 71(4), 787-789.

Van Den Brink, A. M., McLay, C. L., Hosie, A. M., & Dunnington, M. J. (2012). The effect of temperature on brood duration in three *Halicarcinus anger* species (Crustacea: Brachyura: Hymenosomatidae). Journal of the Marine Biological Association of the United Kingdom, 92(3), 515-520.

Van Den Brink, A., Godschalk, M., Smaal, A., Lindeboom, H., & McLay, C. (2013). Some like it hot: the effect of temperature on brood development in the invasive crab *Hemigrapsus takanoi* (Decapoda: Brachyura: Varunidae). Journal of the Marine Biological Association of the United Kingdom, 93(1), 189-196.

Vergés, A., Steinberg, P. D., Hay, M. E., Poore, A. G., Campbell, A. H., Ballesteros, E., Heck, K. L., Booth, D. J., Coleman, M. A., Feary, D. A., Figueira, W., Langlois, T., Marzinelli, E. M., Mizerek, T., Mumby, P. J., Nakamura, Y., Roughan, M., Sebille, E., Sen-Gupta, A., … & Wilson, S. K. (2014). The tropicalization of temperate marine ecosystems: climate-mediated changes in herbivory and community phase shifts. Proceedings of the Royal Society of London, Series B: Biological Sciences, 281(1789), 20140846.

Verza, E., & Cattozzo, L. (2015). Atlante lagunare costiero del Delta del Po. Consorzio di Bonifica Delta del Po, Regione del Veneto, Associazione Culturale Naturalistica Sagittaria.

Vilà, M., Espinar, J. L., Hejda, M., Hulme, P. E., Jarošík, V., Maron, J. L., ... & Pyšek, P. (2011). Ecological impacts of invasive alien plants: a meta-analysis of their effects on species, communities and ecosystems. Ecology Letters, 14(7), 702-708.

Zamora-Marín, J. M., Herrero-Reyes, A. A., Ruiz-Navarro, A., & Oliva-Paterna, F. J. (2023). Non-indigenous aquatic fauna in transitional waters from the Spanish Mediterranean coast: A comprehensive assessment. Marine Pollution Bulletin, 191, 114893.

Zenetos, A., Albano, P. G., Garcia, E. L., Stern, N., Tsiamis, K., & Galanidi, M. (2022). Established non-indigenous species increased by 40% in 11 years in the Mediterranean Sea. Mediterranean Marine Science, 23(1).

Zeng, C. (2007). Induced out-of-season spawning of the mud crab, *Scylla paramamosain* (Estampador) and effects of temperature on embryo development. Aquaculture Research, 38(14), 1478-1485.

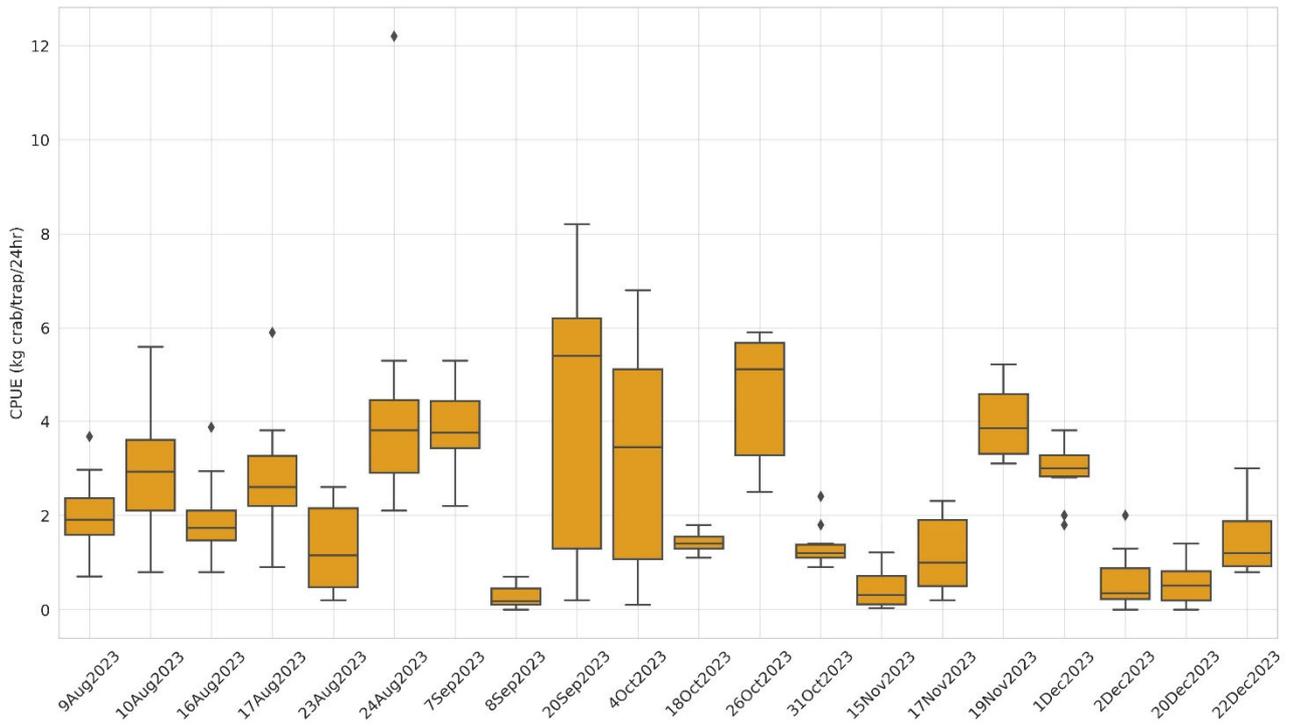

**Figure 1.** *Temporal fluctuations of the traps net yields in the Po Delta area (Italy), shown as catches for unit efforts (hereafter: CPUE) in kg crab/trap/24hr over time.*

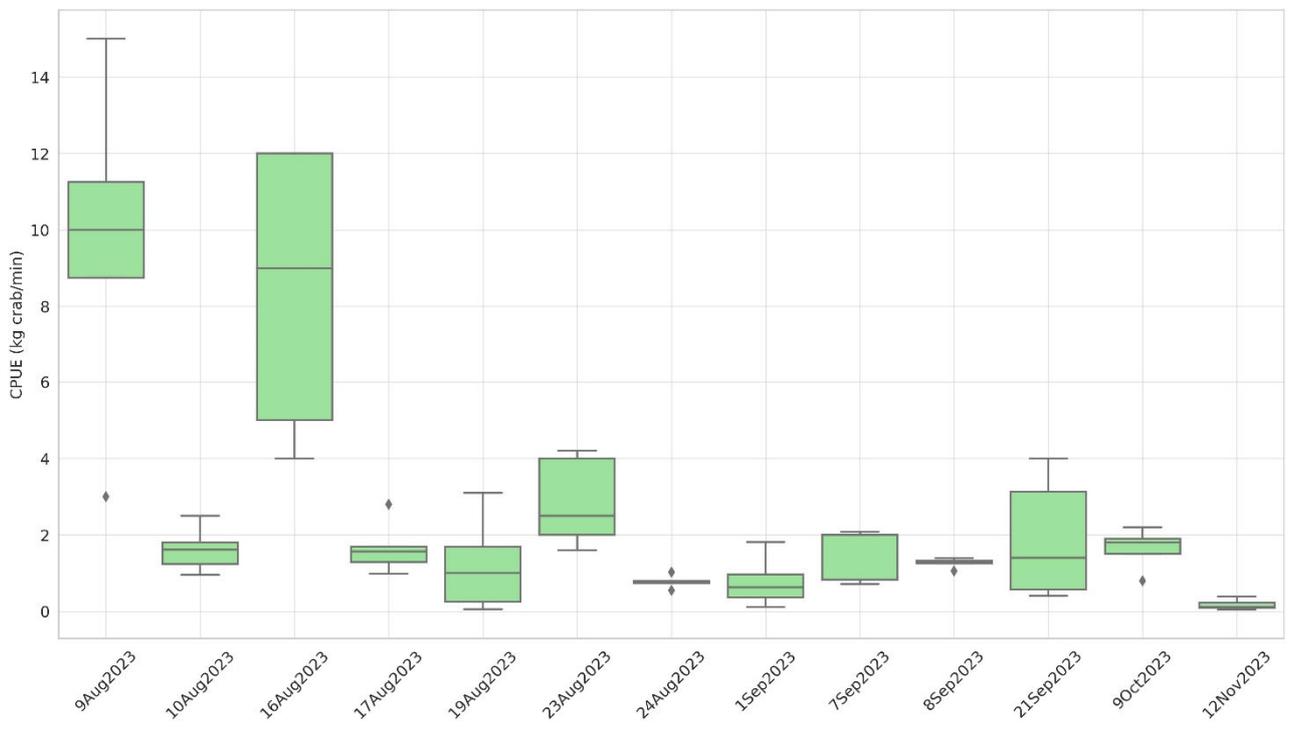

**Figure 2.** *Ostreghero net yields in the Po Delta area (Italy) as CPUE kg crab/min over time.*

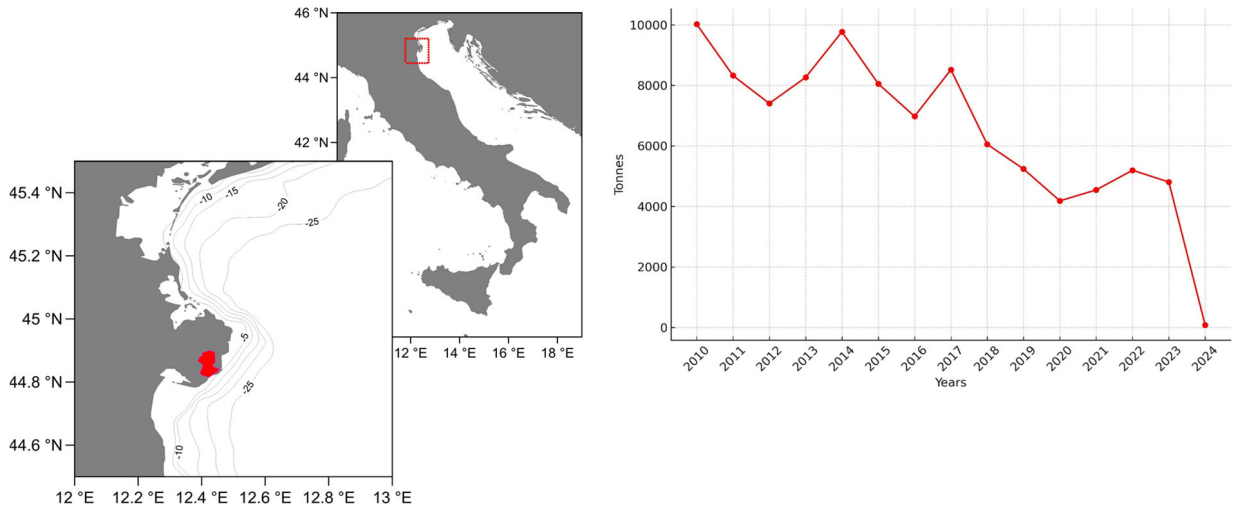

**Figure 3.** *Collapsing clams' production from 2010 to 2024 in the investigated area (marked in red).*

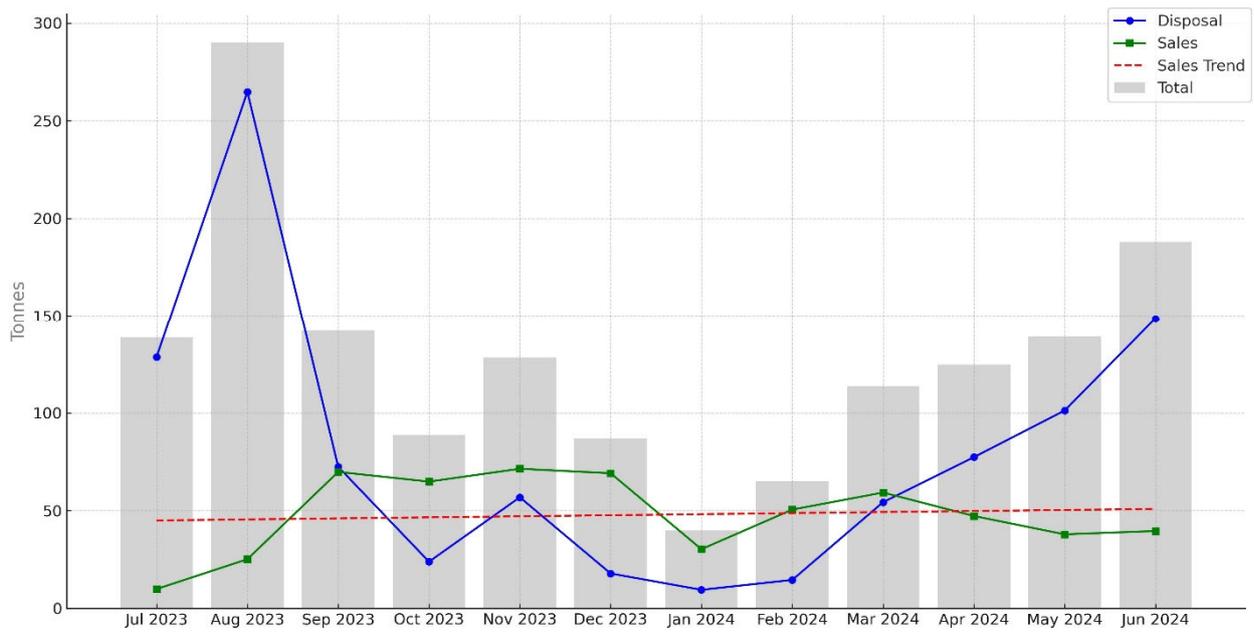

**Figure 4.** *Disposed (blue line) vs. sold (green line) crab, in red the trendline for sales. More explanations in the text.*

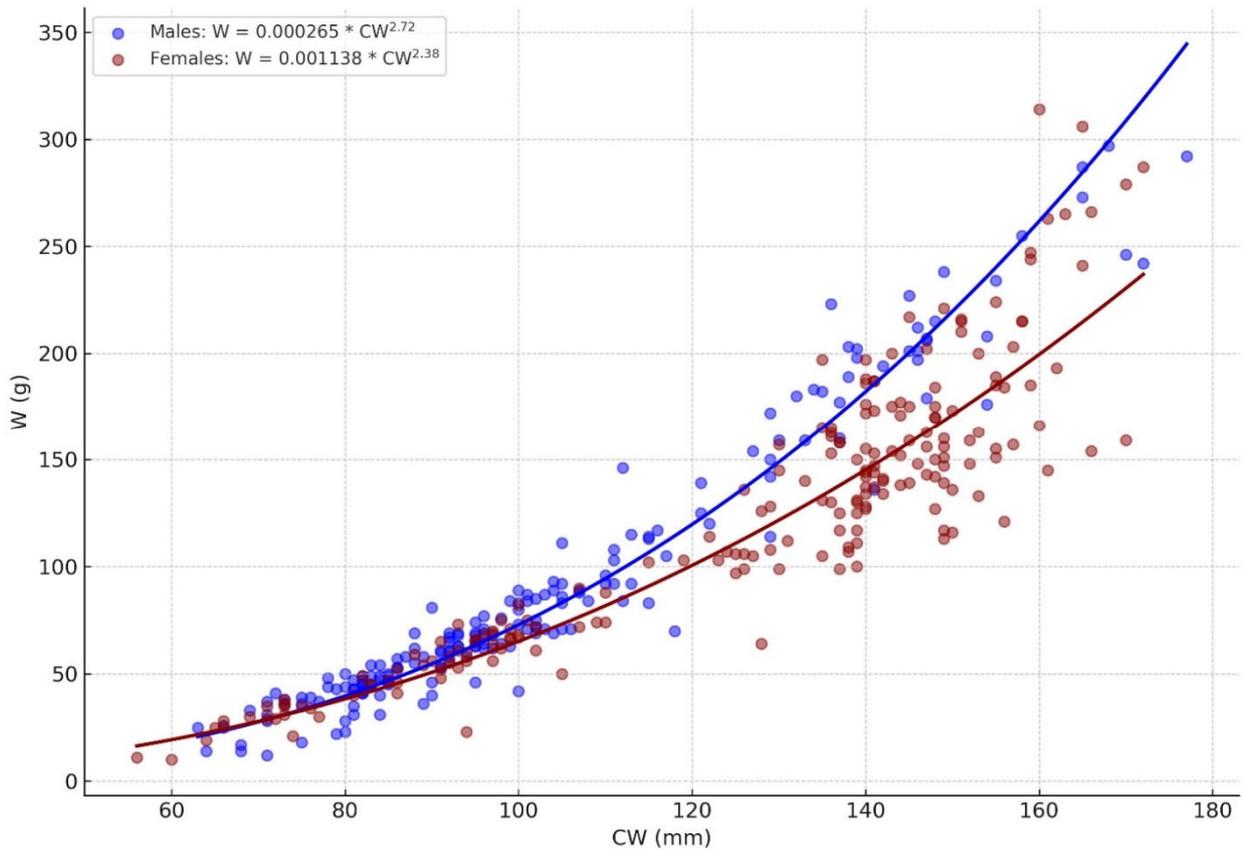

**Figure 5.** *Allometric relationships between the carapace width (CW on the x-axis) and the weight (W on the y-axis) of males and females of* C. sapidus *in the Po Delta area (Italy).*

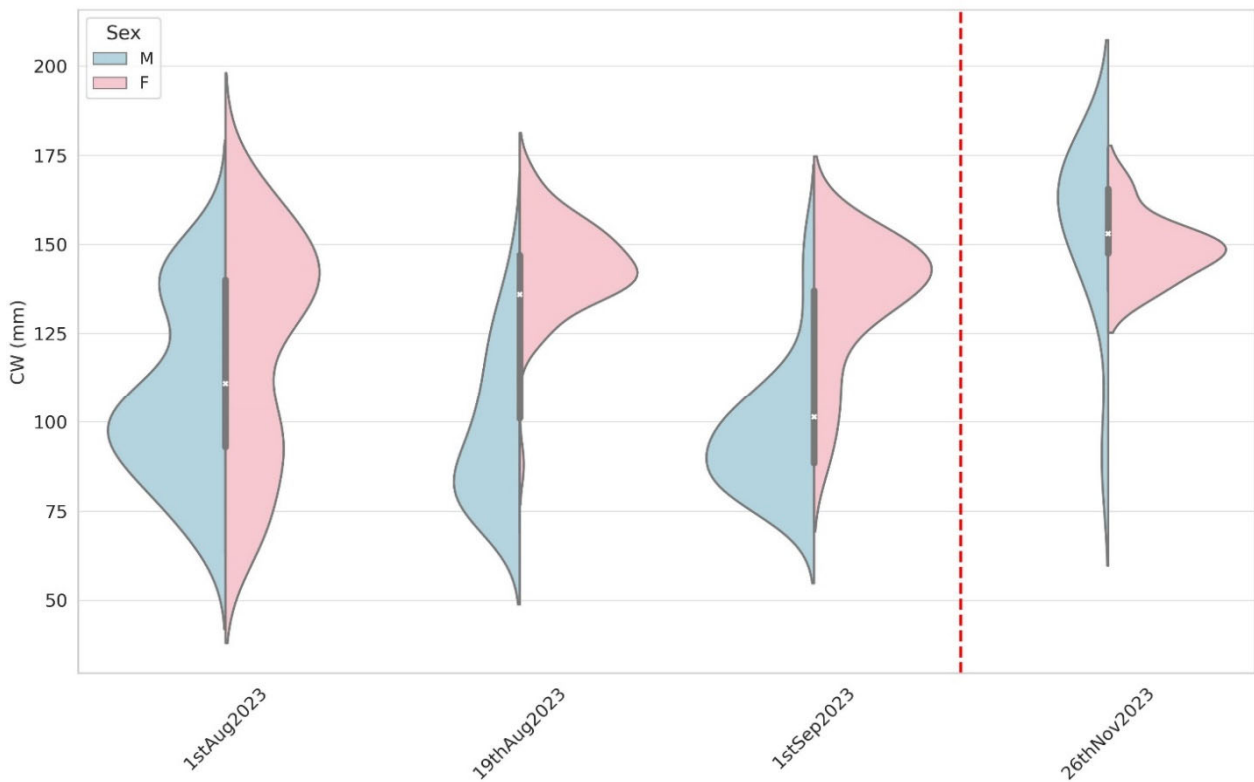

**Figure 6.** *Violin plots with size comparisons (CW) and distribution between male (M) and female (F) specimens over different time periods. The dotted vertical line indicates one clearly abrupt change in the trend between the males/females size distribution. Please see the text for more explanations.*

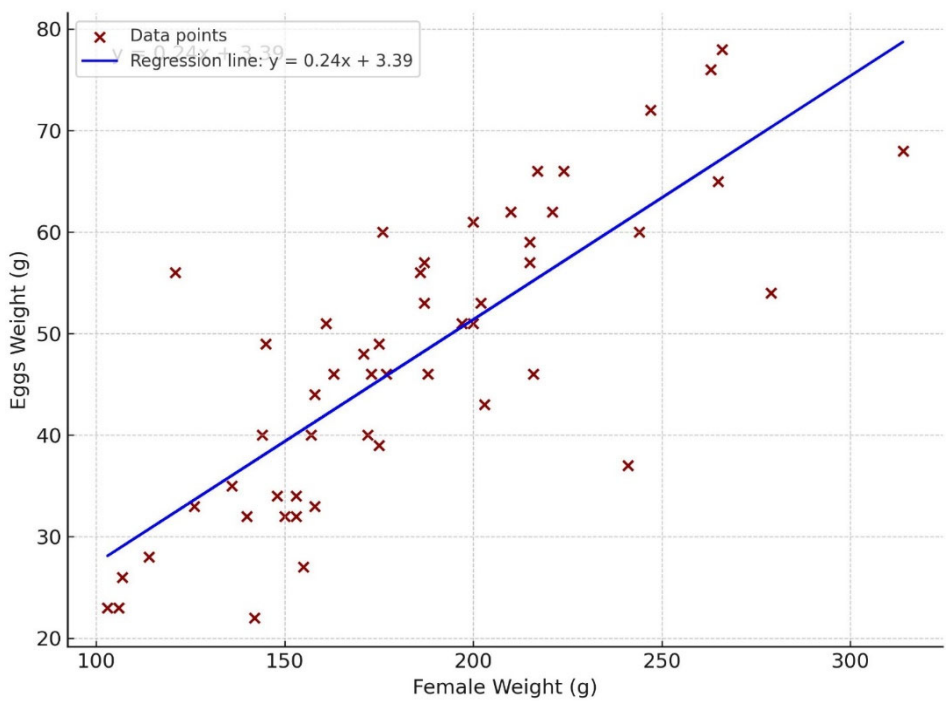

**Figure 7.** *Females weight vs. eggs mass weight of the blue crab (*C. sapidus*) in the Po Delta area.*

**Table 1.** *Biomass of crabs caught with crab traps during sampling period.*

| kg of crabs per period - Crab traps | | | |
| --- | --- | --- | --- |
| Date | Range | Mean | SD |
| 2023-08-09 | 0.5—2.6 | 1.43 | 0.61 |
| 2023-08-10 | 0.8—5.6 | 2.92 | 1.40 |
| 2023-08-16 | 0.6—2.9 | 1.43 | 0.69 |
| 2023-08-17 | 0.9—5.9 | 2.82 | 1.37 |
| 2023-08-23 | 0.2—2.6 | 1.3 | 0.93 |
| 2023-08-24 | 2.1—12.2 | 4.42 | 2.92 |
| 2023-09-07 | 2.2—5.3 | 3.8 | 1.00 |
| 2023-09-08 | 0—0.7 | 0.28 | 0.24 |
| 2023-09-20 | 0.2—8.2 | 4.16 | 2.85 |
| 2023-10-04 | 0.1—6.8 | 3.12 | 2.26 |
| 2023-10-18 | 1.1—1.8 | 1.43 | 0.23 |
| 2023-10-26 | 2.5—5.9 | 4.55 | 1.40 |
| 2023-10-31 | 0.9—2.4 | 1.33 | 0.46 |
| 2023-11-15 | 0.07—2.53 | 0.93 | 0.80 |
| 2023-11-17 | 0.2—2.3 | 1.15 | 0.81 |
| 2023-11-19 | 3.1—5.2 | 3.99 | 0.78 |
| 2023-12-01 | 1.8—3.8 | 2.91 | 0.60 |
| 2023-12-02 | 0—2 | 0.62 | 0.62 |
| 2023-12-20 | 0—1.4 | 0.54 | 0.43 |
| 2023-12-22 | 0.8—3 | 1.46 | 0.70 |

**Table 2.** *Biomass of crabs caught with the "ostreghero net" during sampling period.*

| kg of crabs per period in Min - Ostreghero | | | |
|---|---|---|---|
| Date | Range | Mean | SD |
| 2023-08-09 | 3—15 | 9.60 | 4.37 |
| 2023-08-10 | 0.95—2.5 | 1.62 | 0.59 |
| 2023-08-16 | 4—12 | 8.40 | 3.78 |
| 2023-08-17 | 0.98—2.8 | 1.66 | 0.69 |
| 2023-08-19 | 0.05—3.1 | 1.19 | 1.12 |
| 2023-08-23 | 1.6—4.2 | 2.86 | 1.18 |
| 2023-08-24 | 0.55—1.02 | 0.76 | 0.16 |
| 2023-09-01 | 0.11—1.81 | 0.74 | 0.58 |
| 2023-09-07 | 0.71—2.08 | 1.53 | 0.69 |
| 2023-09-08 | 1.05—1.38 | 1.26 | 0.13 |
| 2023-09-21 | 0.4—4 | 1.82 | 1.38 |
| 2023-10-09 | 0.8—2.2 | 1.64 | 0.53 |
| 2023-11-12 | 0.03—0.38 | 0.17 | 0.13 |

**Table 3.** *Size distribution of* C. sapidus *individuals during sampling period. Abbreviations as follows: carapace width (CW), carapace length (CL), and weight (W).*

| | Combined | | |
|---|---|---|---|
| | Mean | SD | Range |
| **CW** | 114.2 | 31.10 | 30–177 |
| **CL** | 55.49 | 13.29 | 6–166 |
| **W** | 107.8 | 67.15 | 1–314 |
| | Male | | |
| | Mean | SD | Range |
| **CW** | 103.8 | 25.69 | 63–177 |
| **CL** | 53.33 | 11.17 | 33–84 |
| **W** | 93.10 | 65.26 | 12–297 |
| | Female | | |
| | Mean | SD | Range |
| **CW** | 126.1 | 28.91 | 56–172 |
| **CL** | 58.62 | 13.24 | 6–166 |
| **W** | 124.7 | 63.57 | 10–314 |